\documentclass{pasj01}
\Received{$\langle$reception date$\rangle$}
\Accepted{$\langle$acception date$\rangle$}
\Published{$\langle$publication date$\rangle$}
\usepackage{multirow}
\usepackage{comment}
\usepackage{color}

\begin{document}
\title{\textit{NuSTAR} discovery of the hard X-ray emission and a wide-band X-ray spectrum from the Pictor A western hot spot}
\author{%
    Yuji \textsc{Sunada},\altaffilmark{1*}
    Arisa \textsc{Morimoto},\altaffilmark{1}
    Makoto S. \textsc{Tashiro},\altaffilmark{1,2}
    Yukikatsu \textsc{Terada},\altaffilmark{1,2}
    Satoru \textsc{Katsuda},\altaffilmark{1}
    Kosuke \textsc{Sato},\altaffilmark{1}
    Dai \textsc{Tateishi},\altaffilmark{1} and
    Nobuaki \textsc{Sasaki}\altaffilmark{1}
}
\altaffiltext{1}{Graduate School of Science and Engineering, Saitama University, Shimo-Okubo 255, Sakura, Saitama 338-8570}
\altaffiltext{2}{Institute of Space and Astronautical Science (ISAS), Japan Aerospace Exploration Agency (JAXA), 3-1-1 Yoshinodai, Chuo-ku, Sagamihara, Kanagawa, 252-5210, Japan}
\email{sunada@heal.phy.saitama-u.ac.jp}


\maketitle
\begin{abstract}
Utilizing \textit{Chandra}, \textit{XMM-Newton} and \textit{NuSTAR}, a wide-band X-ray spectrum through 0.2 to 20 keV is reported from the western hot spot of Pictor A.
In particular, 
the X-ray emission is significantly detected in the 3 to 20 keV band at 30 sigma by \textit{NuSTAR}.
This is the first detection of hard X-rays with energies above 10 keV from a jet termination hot spot of active galactic nuclei.
The hard X-ray spectrum is well described with a power-law model with a photon index of $\mathit{\Gamma}=1.8\pm0.2$, and the flux is obtained to be $(4.5\pm0.4)\times10^{-13}$ erg s$^{-1}$ cm$^{-2}$ in the 3 to 20 keV band.
The obtained spectrum is smoothly connected with those soft X-ray spectra observed by \textit{Chandra} and \textit{XMM-Newton}.
The wide-band spectrum shows a single power-law spectrum with a photon index of $\mathit{\Gamma}=2.07\pm0.03$, excluding any cut-off/break features.
Assuming the X-rays as synchrotron radiation of the electrons, the energy index of the electrons is estimated as $p=2\mathit{\Gamma}-1=3.14\pm0.06$ from the wide-band spectrum.
Given that the X-ray synchrotron emitting electrons quickly lose their initial energies via synchrotron radiation, the energy index of electrons at acceleration sites is estimated as $p_\mathrm{acc}=p-1=2.14\pm0.06$.
This is consistent with the prediction of the diffusive shock acceleration.
Since the spectrum has no cut-off feature up to 20 keV, the maximum electron energy is estimated to be no less than 40 TeV.
\end{abstract}

\clearpage
\section{Introduction}
Bright ``hot spots" observed at the terminal of the relativistic jets in Fanaroff-Riley type-II (FR-II, \cite{fanaroff74}) radio galaxies are the locations where the active galactic nucleus jet converts their kinetic energy into accelerated particles.
Their emissions from radio to near-infrared/optical band are explained by the synchrotron radiation from the accelerated electrons ( e.g., \cite{meisenheimer89,meisenheimer97,mack09,Werner12}). 
The X-ray components brighter and harder than the extrapolation of the low-energy ones (e.g., \cite{harris94,wilson00}) are often explained by  the one-zone Synchrotron-Self-Compton (SSC, \cite{jones74,band85}) for radio-loud hot spots, such as those in the Cygnus A (e.g., \cite{harris00,hardcastle04,kataoka05,stawarz07,Werner12}).
On the other hand, for the radio-quiet ones, the SSC X-ray scenario is unlikely since it requires magnetic field strength so low (i.e., $B \lesssim 10 \ \mathrm{\mu G}$, \cite{zhang18}) that it is far below the value expected in equipartition condition (i.e., $B \sim 100 \ \mathrm{\mu G}$, \cite{kataoka05}).
Moreover, some of these sources indicate spatial discrepancy between the radio and X-ray emission regions (e.g.,\cite{thimmappa20,orienti20,migliori20}).
These features imply multi-zone/component radiations, such as the synchrotron radiation of another electron population is considered plausible X-ray origin (e.g., \cite{hardcastle04}).

The X-ray synchrotron emission corresponds to the electrons with energies of a few tens of TeV, and the derived radiative lifetime of the electrons is an order of 10 years.
This means that the X-ray, particularly in hard X-ray, reflects the recent particle acceleration.
Thus, X-ray observations are valuable to study the accelerated particles and the recent history of particle acceleration in hot spots.
However, in general, a high angular resolution is crucial to detect hot spots by resolving them from the nuclei.
In addition, a wide-band X-ray spectrum is required to determine the energy index and investigate possible cut-off/break features, which reveals the physical state in the hot spot.
Most hot spots are detected with limited X-ray energy range and photon statistics, and then a detailed spectrum is not well known.

The hot spot located at the western jet terminal of Pictor A is one of the nearest and ideal hot spots to be investigated in X-rays.
The hot spot located 4 arcmin from the nucleus exhibits one of the highest X-ray flux among the hot spots (see the compilations of \cite{kataoka05} and \cite{zhang18}).
So far, some remarkable features are reported in multi-wavelength \citep{thomson95,wilson01,isobe17,isobe20,thimmappa20}.
\citet{tingay08} discovered sub-kpc-scale radio structures in the hot spot and suggested them to be counterparts of the X-ray emission.
\citet{hardcastle16} reported the possible month-scale X-ray flux decrease from the long-term \textit{Chandra} observation and supported idea that the fine radio structure is physically associated with the X-ray emission because of the compatibility between the time scale and the size of the fine structure.
In addition to the flux decrease, they reported a possible spectral break around 2 keV.
The flux decrease and spectral break led them to argue that X-rays are emitted by the highest-energy electron population.
However, there remains room for further investigation due to the limited statistics and energy range.

The high-energy cut-off is predicted in X-ray bands if the X-rays originate from the highest energy electrons.
In this paper, we utilize the hard X-ray observation performed by \textit{NuSTAR} to investigate the cut-off feature.
Ahead of \textit{NuSTAR} data analysis, we present \textit{Chandra} and \textit{XMM-Newton} data analysis to investigate the wide-band X-ray spectral shape with high photon statistic and verify the flux decrease.

Throughout this paper, we adopted the cosmological parameters of $H_0=67$ km s$^{-1}$ Mpc$^{-1}$, $\Omega_\mathrm{m}=0.32$, $\Omega_\Lambda=0.68$ \citep{planck20}.
At the redshift of Pictor A ($z=0.035$, \cite{eracleous04}), the luminosity distance is evaluated as $158$ Mpc.
The angular size of 1 arcsec corresponds to the physical size of 0.71 kpc at the source rest frame.

\section{Observation and data reduction}

\subsection{\textit{Chandra}}
\textit{Chandra} \citep{weisskopf00} performs the high resolution soft X-ray imaging and spectroscopy.
The High Resolution Mirror Assembly of \textit{Chandra}, realizing a sub-arcsec scale PSF size, is the best X-ray optics for us to investigate a detailed position and structure of the hot spot.
The X-ray detectors of Advanced charge-coupled device (CCD) Imaging Spectrometer (ACIS: \cite{garmire03}) cover the energy range of 0.3--10.0 keV.

\textit{Chandra} observed Pictor A region 14 times between 2000 and 2015.
The observation IDs and the dates are tabulated in Table \ref{tbl:ChandraObs}.
These observation results are reported in papers.
The X-ray hot spot was found to be well associated with the radio one and identified by the \textit{Chandra} \citep{wilson01,thimmappa20}.
Furthermore, \citet{hardcastle16} investigated all observations and reported an X-ray flux decrease of the hot spot.

We re-analyzed the data of the observations to confirm the X-ray emission from the hot spot and re-evaluate the time variability, utilizing the software (\texttt{ciao-4.13}, \cite{fruscione06}) and the calibration database (\texttt{CALDB 4.9.5}), which are newer version than those used in the literature.
We performed the data reduction using \texttt{ chandra\_repro} script.
The obtained exposure times after the reduction are almost the same as those in \citet{hardcastle16}, a total of which is 460 ksec.

\subsection{\textit{XMM-Newton}}
\textit{XMM-Newton} \citep{jansen01} performs the  high throughput imaging spectroscopy in a soft X-ray band with the three telescopes realizing a large effective area of 4,500 cm$^2$ at 1 keV, in total.
Their field of view of the size of $30\times30$ arcmin$^2$ and PSF size of $\sim$ 15 arcsec in Half Power Diameter (HPD) are well suited to the hot spot observation.
Three X-ray CCD cameras (European Photon Imaging Camerras: EPICs) installed in the telescopes, two front-illuminated CCDs of Metal Oxide Semi-conductors (MOS1,2 \cite{turner01}) and a back-illuminated pn type CCD (PN \cite{struder01}) cover the energy range of 0.2--10 keV.

\textit{XMM-Newton} observed the Pictor A region on 2001 March 17 and on 2005 January 14.
The observation IDs are 0090050801 and 0206039010, respectively.
Utilizing the data, \citet{grandi03} and \citet{migliori07} studied the eastern and the western lobes of Pictor A.

In the first observation, the MOS cameras were operated in the small window mode and failed to observe the western hot spot.
The PN observed the hot spot, but near the detector gap.
In the second observation, on the other hand, the hot spot was successfully observed by both MOS1 and PN, but was out of the field of view of MOS2 as it was operated in the small window mode.
Therefore we analyzed the MOS1 and PN data obtained in the second observation.
We performed the data reduction using the Science Analysis Software (SAS) version 19.0.0.
The effective exposure times are 49 and 30 ksec for MOS1 and PN, respectively.

\subsection{\textit{NuSTAR}} 
The \textit{NuSTAR} satellite \citep{harrison13} realizes hard X-ray imaging spectroscopy.
The equipped two co-aligned Walter-type I telescopes have almost the same field of view of the size of $13\times13$ arcmin$^{2}$, with an excellent PSF of 58 arcsec in HPD.
\textit{NuSTAR} observes both the nucleus and the hot spot in a field of view and resolves them.
The Focal Plane Module A and B (FPMA and B) covers the energy range of 3-79 keV with the pixelized CdZnTe detector arrays.

In this work, we utilized the archival data aiming at the  Pictor A nucleus (obs.ID 60101047002).
The observation was carried out between December 3 and 6 in 2015.
For the \textit{NuSTAR} data reduction and analysis, we used the software of NUSTARDAS ver. 2.0.0 and calibration database of CALDB ver 20210524.
We performed the data reduction using \texttt{nupipeline} script with standard criteria for science observation named ``SCIENCE.''
The resultant effective exposure is 109 ksec.

\section{Analysis and Result}
\subsection{Image analysis}
Before analyzing the \textit{NuSTAR} and \textit{XMM-Newton} images which are performed on the hot spot in this paper for the first time, we reviewed the spatial property of the hot spot on the \textit{Chandra} image.
We employed the observation of Obs.ID 4369, whose aim point is located at the hot spot of our interest.
The X-ray hot spot is clearly detected and is found to be slightly extended by 5 arcsec, with a tiny central region providing the major contribution of the flux.  
In fact, the size of the hot spot is measured to be 1.5 arcsec in HPD, which, we assume, corresponds to the `compact' component named in \citet{hardcastle16}.
The morphology is consistent with the literatures \citep{hardcastle16,thimmappa20}.

\subsubsection{\textit{XMM-Newton} image}
We analyzed the images of \textit{XMM-Newton}/MOS1 and PN.
We confirmed the point-like sources associated with the hot spot as mentioned in \citet{grandi03} and \citet{migliori07}.
Its position is consistent with the hot spot within 2-sigma level pointing accuracy of 3 arcsec (MOS1) and 2 arcsec (PN) \citep{kirsch04}.
We found no significant X-ray extension beyond the PSF, which is reasonable given the size determined by \textit{Chandra}.
There is no confusing source in \textit{Chandra} image. 
Therefore we can safely conclude that the source detected with \textit{XMM-Newton} is the hot spot itself.

\subsubsection{\textit{NuSTAR} image}
\begin{figure*}
 \begin{center}
  \includegraphics[width=\linewidth,bb= 0 0 247 91]{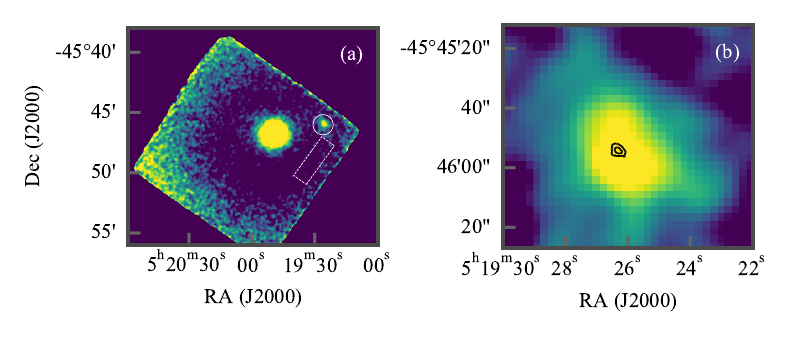}
 \end{center}
 \vspace{-5mm}
 \caption{(a) FPMA exposure-corrected image smoothed with a Gaussian with the standard deviation of 4.8 arcsec. The dashed circle and rectangular indicate source and background regions, respectively. (b) The enlarged image around the western region overlaid the contour of \textit{Chandra} image is shown. }\label{fig:FPM_images}
\end{figure*}

We show both FPMA and FPMB exposure-corrected images and found two sources as shown in panel (a) of Figure \ref{fig:FPM_images}.
The exposure map was generated by the \texttt{nuexpomap} script.
The image was smoothed by Gaussian with a standard deviation of 4.8 arcsec, corresponding to twice pixel size, for the visibility.

The brighter source at the center of the image is Pictor A nucleus, whose detailed spectral analysis was already reported in \citet{kang20}.
In addition to the nucleus, we see another X-ray source in the northwest.
As shown in the enlarged image in panel (b) of Figure \ref{fig:FPM_images}, with overlaid contours of \textit{Chandra} image, we see that the hard X-ray source is well associated with the hot spot.
The significance of this hard X-ray emission with \textit{NuSTAR} was measured to be 30 $\sigma$ from the statistical 
error of the background. 

To investigate the candidate hard X-ray emission of the hot spot, we evaluated the spatial properties quantitatively.
We found that positions of the brightest pixels in FPMA and FPMB are 2.5 and 2.8 arcsec away from the position determined by \textit{Chandra}, respectively.
These are within the pointing uncertainty of the \textit{NuSTAR} of 8 arcsec \citep{harrison13}.

The apparent size is measured on the un-smoothed image (i.e., the raw image of Figure \ref{fig:FPM_images} b) as $35$ arcsec in HPD, 
which is smaller than the \textit{NuSTAR} PSF size of 58 arcsec in HPD.
Therefore, the major emission from the object is considered to be a point-like object for \textit{NuSTAR}.
Since the evaluated soft X-ray hot spot size is negligible compared to the PSF of 
\textit{NuSTAR}, the obtained image size is consistent with the hot spot.
In addition, there is no other bright X-ray source around the hot spot in \textit{Chandra} image as mentioned above.
Therefore, we safely concluded that the observed source is a hard X-ray counterpart of the hot spot.
This is the first time that a jet termination hot spot has been detected in the energy range above 10 keV.

\subsection{Time variability analysis using \textit{Chandra} data}
\label{sec:chandra_time_val}
\citet{hardcastle16} reported an interesting X-ray behavior --- a month-scale flux decrease of the hot spot. 
They suggested that it could be caused by energy losses of the highest energy electrons.
However, the flux decrease is accompanied by a spectral hardening, which may imply a flux decrease in the low-energy band (see, \cite{hardcastle16,thimmappa20}).
Here, we carefully examined the unexpected increase of molecular contamination on the detector surface, which reduces the effective area below 1 keV band.
In fact, a position dependence of the contamination is reported, although the time variation at the typical aim point is calibrated and corrected with the CALDB file \citep{plucinsky18}.

\begin{table}
  \tbl{Observation IDs and dates of the \textit{Chandra} observations}
      {
        \begin{tabular}{ccc|ccc}
          \hline
          ID& date& Epoch& ID& date& Epoch\\
          \hline
          346 & 2000-01-18 & 1 & 14221 & 2012-11-06 & 5\\
          3090 & 2002-09-17 & 2 & 15580 & 2012-11-08 & 5\\
          4369 & 2002-09-22 & 2 & 15593 & 2013-08-23 & 6\\
          12039 & 2009-12-07 & 3 & 14222 & 2014-01-17 & 7\\ 
          12040 & 2009-12-09 & 3 & 14223 & 2014-04-21 & 8\\ 
          11586 & 2009-12-12 & 3 & 16478 & 2015-01-09 & 9\\
          14357 & 2012-06-17 & 4 & 17574 & 2015-01-10 & 9\\
            \hline
        \end{tabular}
      }
      \label{tbl:ChandraObs}
\end{table}

\begin{figure}
 \begin{center}
  \includegraphics[width=\linewidth,bb=0 0 238 241]{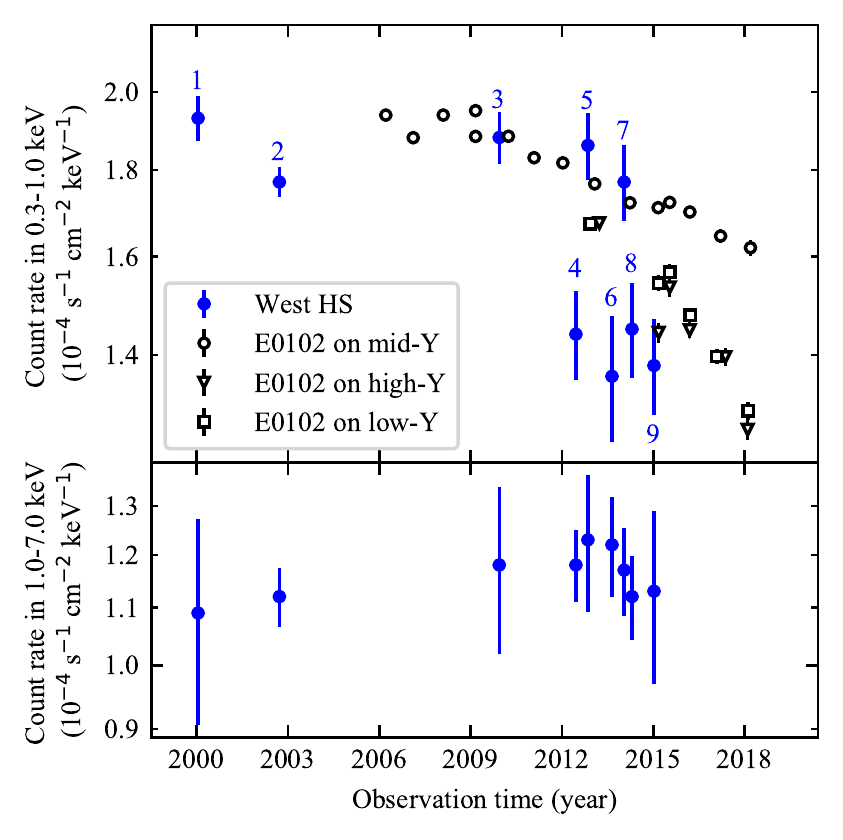}
 \end{center}
 \caption{The effective area corrected count rates of the hot spot and E0102 are shown.
The count rates of E0102 are scaled for the comparison to the hot spot.
 The blue filled circles indicate the count rates of hot spot and accompanying numbers indicates a corresponding epoch.
 The open circles, open triangles and open squares indicate the E0102 at the mid, high and low-Y positions.
 The upper panel is the low energy band of 0.3\--1.0 keV. The lower panel is the high energy band of 1.0\--7.0 keV.
 }\label{fig:lightcurve_Chandra}
\end{figure}

We investigated the time variation of the count rate of the hot spot in low-energy (0.3--1.0 keV), which is sensitive to contamination, and high-energy (1.0--7.0 keV) bands for comparison separately.
We divided the observation data into 9 epochs, according to \citet{hardcastle16} to show the time history of count rates in Figure \ref{fig:lightcurve_Chandra}.
In the low energy band, the count rates in epochs 4,6,8, and 9 are around 20 per cent lower than those in the other epochs, while there is no significant fluctuation in the high energy band.

\begin{figure}
 \begin{center}
  \includegraphics[width=\linewidth,bb=0 0 270 270]{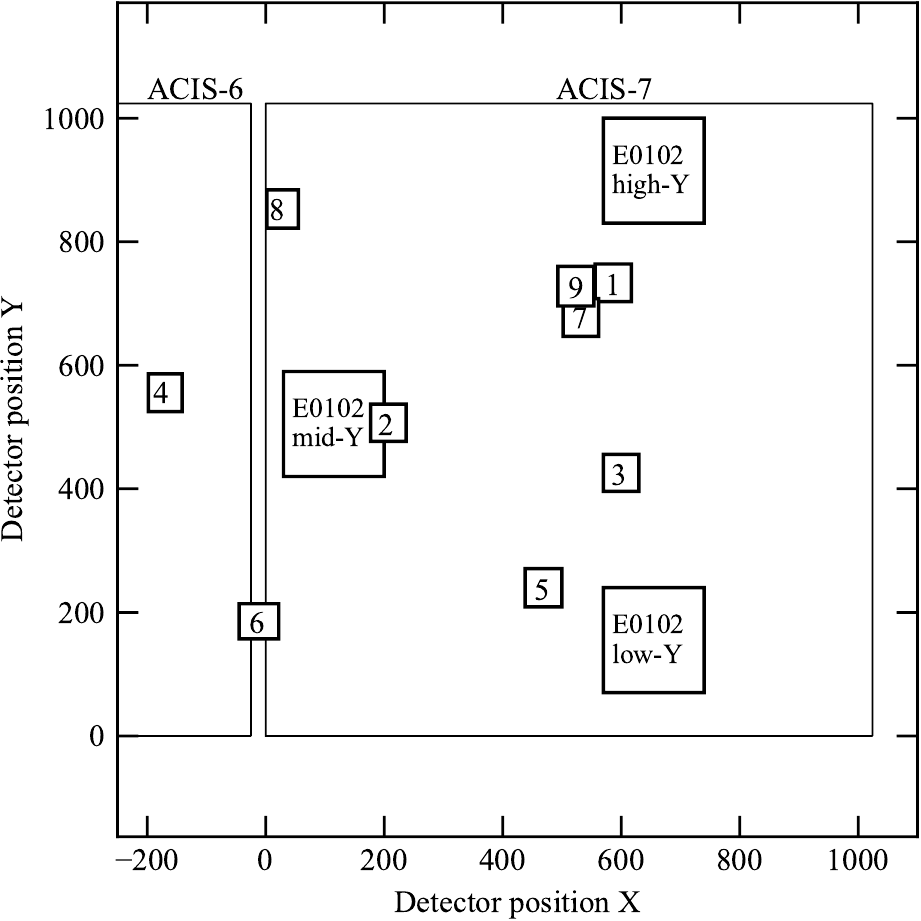}
 \end{center}
 \caption{The hot spot and E0102 detected positions on the ACIS are shown. The small squares surrounding the numbers indicates the hot spot position in each epoch.
 The three middle squares indicate the E0102 positions of low, mid and high-Y.
 The large squares of thin solid line indicates the ACIS chips 6 and 7.}\label{fig:WS_pos_Chandra}
\end{figure}

Figure \ref{fig:WS_pos_Chandra} shows the source position on the detectors for each epoch.
In epochs 4, 6, and 8, the hot spot is close to the chip gap or observed by a different ACIS chip. 
In these cases, we cannot ignore the systematic uncertainties.
To examine the apparent decrease in epoch 9, we also investigated the count rate variation of a calibration source, the supernova remnant E0102, during this period.
E0102 has been observed at three different locations on the detector, indicated as low-Y, mid-Y, and high-Y in Figure \ref{fig:WS_pos_Chandra}.
The obtained count rates are also shown in Figures \ref{fig:lightcurve_Chandra}, respectively.
We see the count rates on high- and low-Y positions have decreased since 2015.
This trend implies that the contamination effect causes the apparent decrease of the count rates on low- and high-Y positions.
The hot spot position in epoch 9 is near the high-Y location.
Thus, we regard the apparent decrease in epoch 9 likely caused by artifacts (building-up contaminants).
The hot spot's location at epoch 8 is not exactly the same as that for E0102, but is located at the high-Y position where rapid increase of contaminants has been suggested. 
Therefore, the count-rate drop observed at epoch 8 would be caused not only by the effects of the chip gap but also by the possible contaminants.
In this context, we argue that the flux decrease after epoch 8 might be affected by the detector response, particularlly for the contaminants on the detector, although our argument does not completely exclude the flux variability reported in \citet{hardcastle16}.
It is important to carefully evaluate the contamination build-up, for more detailed investigations on the time variability.
The results from the X-ray all-sky survey performed by \textit{eROSITA} \citep{eROSITA21} will provide us with an excellent opportunity for this purpose.

\subsection{Spectral analysis}
\subsubsection{\textit{XMM-Newton} spectrum}
We investigated the spectrum obtained from \textit{XMM-Newton}'s second observation of Pictor A.
The source spectrum was extracted from a circular region with a radius of 18 arcsec centered at the brightness peak of the hot spot.
The background spectrum was extracted from a nearby, source-free circular region with a radius of 36 arcsec, in which point sources detected by \texttt{edetect\_chain} script are excluded. 
We adopted the same procedure of spectral extraction for MOS1 and PN.
The source emission significantly exceeds the background one in the energy range of 0.2--10.0 keV in both MOS1 and PN.
The spectral analysis hereafter were performed with \texttt{XSPEC} \citep{arnaud96} version 12.11.1 in \texttt{HEAsoft} package version 6.28.

\begin{figure}
 \begin{center}
  \includegraphics[width=\linewidth,bb=0 0 480 300]{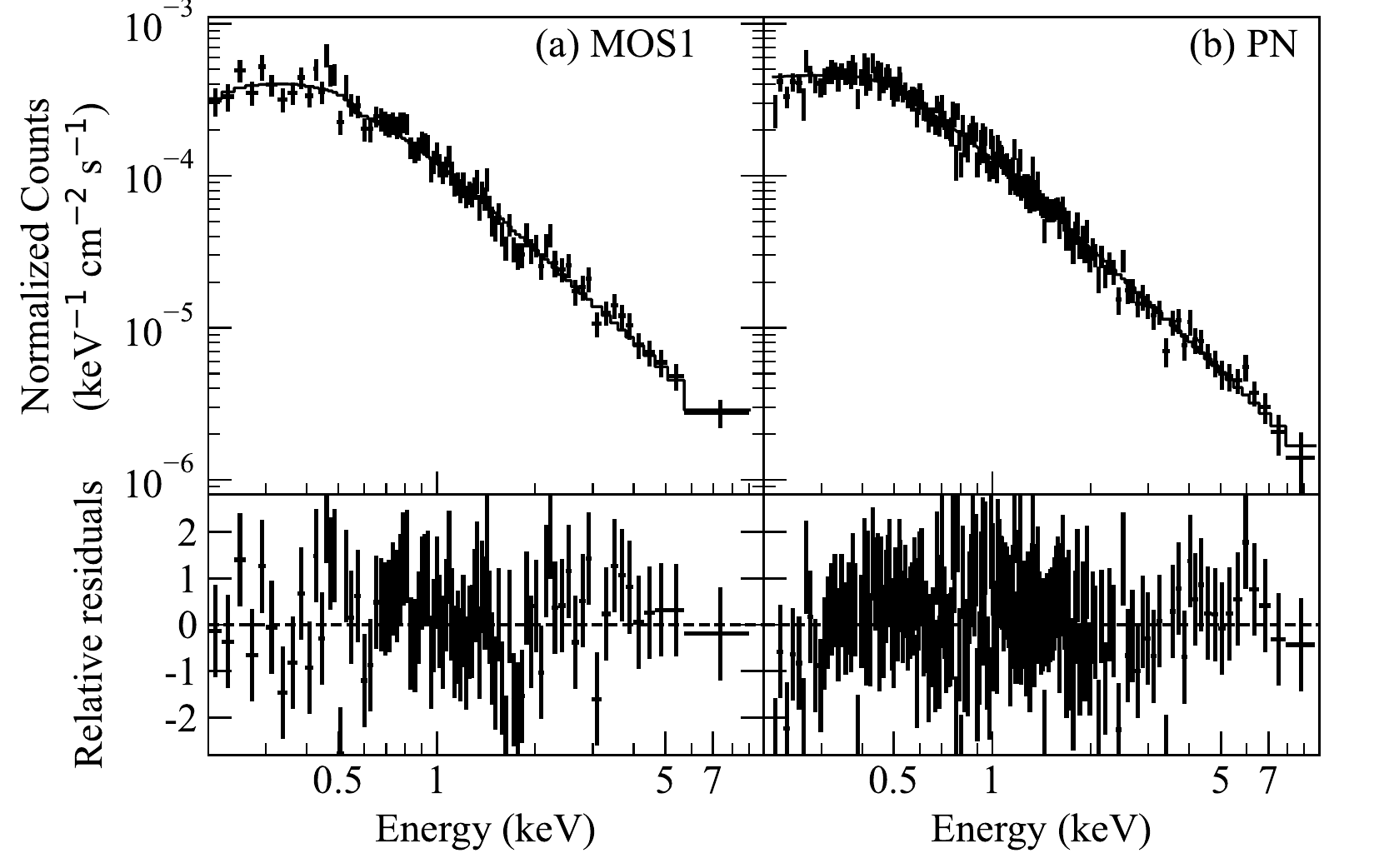}
 \end{center}
 \caption{ The background subtracted spectra of the \textit{XMM-Newton} detectors and best-fit model are shown as the crosses and solid line, respectively. The right and left panels indicate MOS1 and PN, respectively}\label{fitted_spc_xmm}
\end{figure}

We fitted the  background-subtracted spectrum with an absorbed power-law model, described as \texttt{tbabs*pegpwrlw} in \texttt{XSPEC}, using $\chi^2$ statistics.
In the fitting, we fixed the hydrogen column density to the Galactic value of $N_\mathrm{H}$=3.6$\times10^{20}$ atoms cm$^{-2}$ obtained from HI4PI\footnote{The $N_\mathrm{H}$ value is obtained in the website at https://heasarc.gsfc.nasa.gov/cgi-bin/Tools/w3nh/w3nh.pl}\citep{hi4pi16}.
The solar abundance table we used in \texttt{XSPEC} was obtained from \citet{anders89}.
Figure \ref{fitted_spc_xmm} shows the spectra and the best-fit models, and the best-fit parameters and fitting range are summarized in Table \ref{tbl:Fit_Par}.
The models well reproduce the spectra with $\chi^2$/d.o.f. = 97/90 and 197/181 for spectra obtained by MOS1 and PN, respectively.

\begin{figure}
 \begin{center}
  \includegraphics[width=80mm,bb=0 0 240 240]{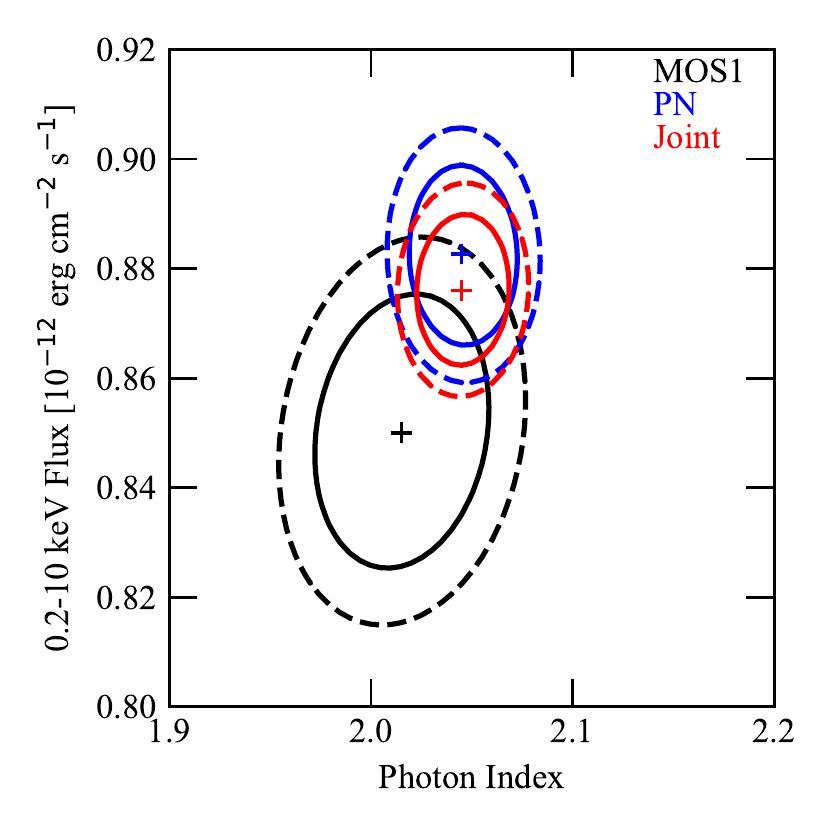}
 \end{center}
 \caption{The confidence contours for the 0.5-7 keV flux and photon index of the MOS1(black), PN(blue) and MOS1 and PN jointed(Red) spectrum fitting. The solid and dashed lines indicate 68\% and 90\% confidence level, respectively. The crosses indicate best-fit parameters.}\label{fig:conturXMM}
\end{figure}

We compared the confidence contours, to check the consistency between the MOS1 and PN spectra.
As shown in Figure \ref{fig:conturXMM}, the contours are consistent with each other at a 68\% confidence level.
Then, we fitted the MOS1 and PN spectra simultaneously.
The best-fit parameters and the confidence contours are shown in Figure \ref{fig:conturXMM} and Table \ref{tbl:Fit_Par}, respectively. 
There is no significant difference between the joint and individual fittings.

\subsubsection{\textit{Chandra} Spectrum}
To improve the photon statistic, we utilized the \textit{Chandra} spectra.
We adopted the circular region with a radius of 8 arcsec and the annulus region with the inner and outer radii of 12 and 20 arcsec for the source and background regions, respectively.
We used the spectrum above 1.5 keV, to avoid the possible contamination effect. 
For a simplification, we combined the spectra in epoch 1--9 using the \texttt{addascaspec} command.
We fitted the combined spectrum with the absorbed power-law model as same as for the \textit{XMM-Newton} spectrum.
The model well reproduced the spectrum with the best-fit chi-square statistics of $\chi^2/$d.o.f.=226/235.
The best-fit parameters are summarized in Table \ref{tbl:Fit_Par}.
The derived parameters are consistent with the analysis of early \textit{Chandra} data in \citet{wilson01}.

\subsubsection{\textit{NuSTAR} Spectrum}
\begin{figure}
 \begin{center}
  \includegraphics[width=\linewidth,bb=0 0 480 300]{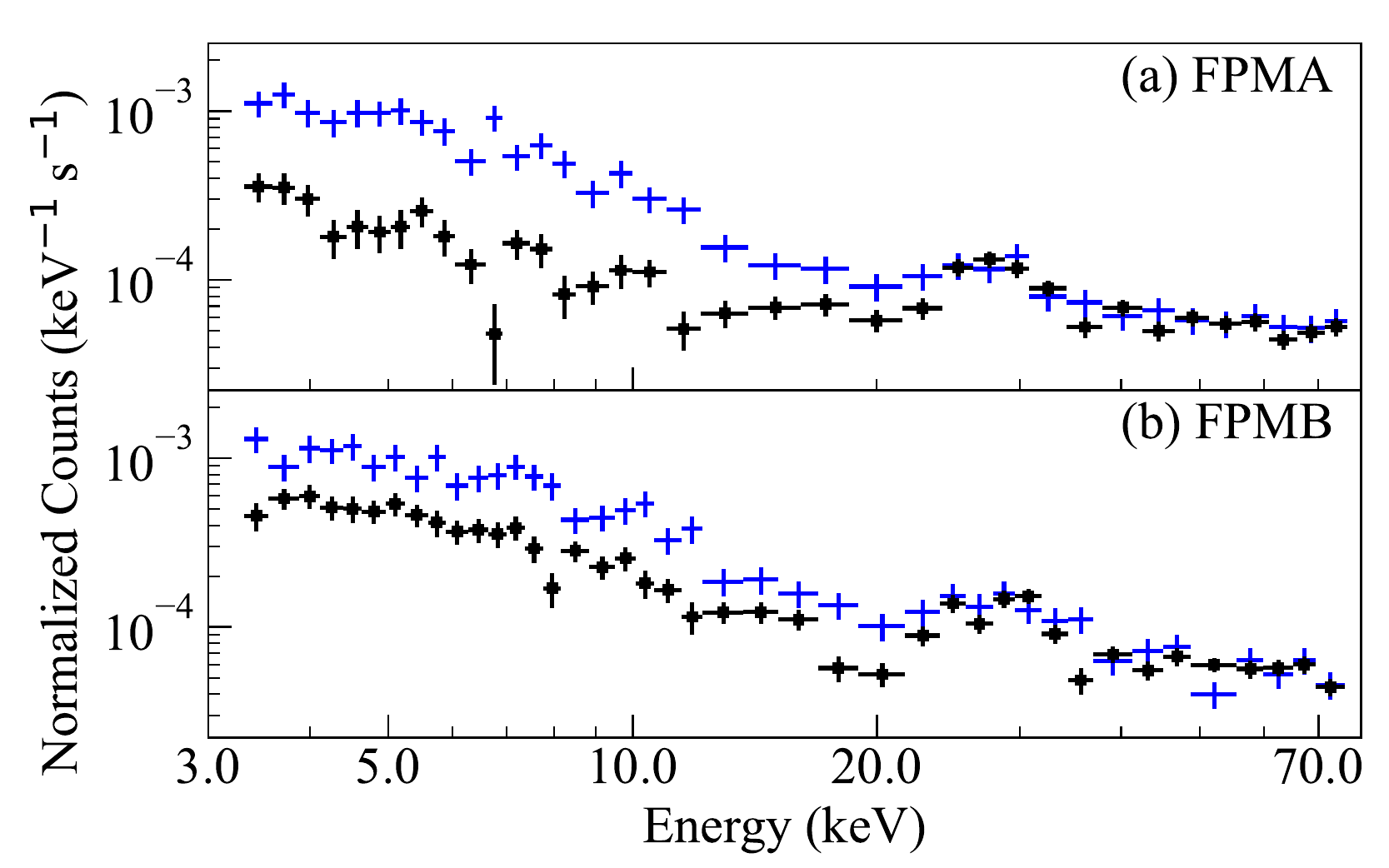}
 \end{center}
 \caption{The source and background spectra are shown as the blue crosses and black squares, respectively. The upper and lower panels indicate FPMA and FPMB, respectively.}\label{src_bkg_spec}
\end{figure}

To investigate the hard X-ray property of the western hot spot, we analyzed the \textit{NuSTAR} spectrum.
As shown in Figure \ref{fig:FPM_images}, we employed a circular region with a radius of 50 arcsec centered at the centroid of the hard X-ray brightness distribution of the western hot spot and a rectangular area in the same detector chip for the source and background, respectively.
We extracted the source and background spectra via the  \texttt{nuproducts} script, which also generates the response matrix functions and the auxiliary response files.
Figure \ref{src_bkg_spec} shows the source and background spectra of the FPMA and FPMB.
We see significant emissions from the hot spot in the energy range of 3--20 keV, in both detectors. 

\begin{figure}
 \begin{center}
  \includegraphics[width=\linewidth,bb=0 0 480 300]{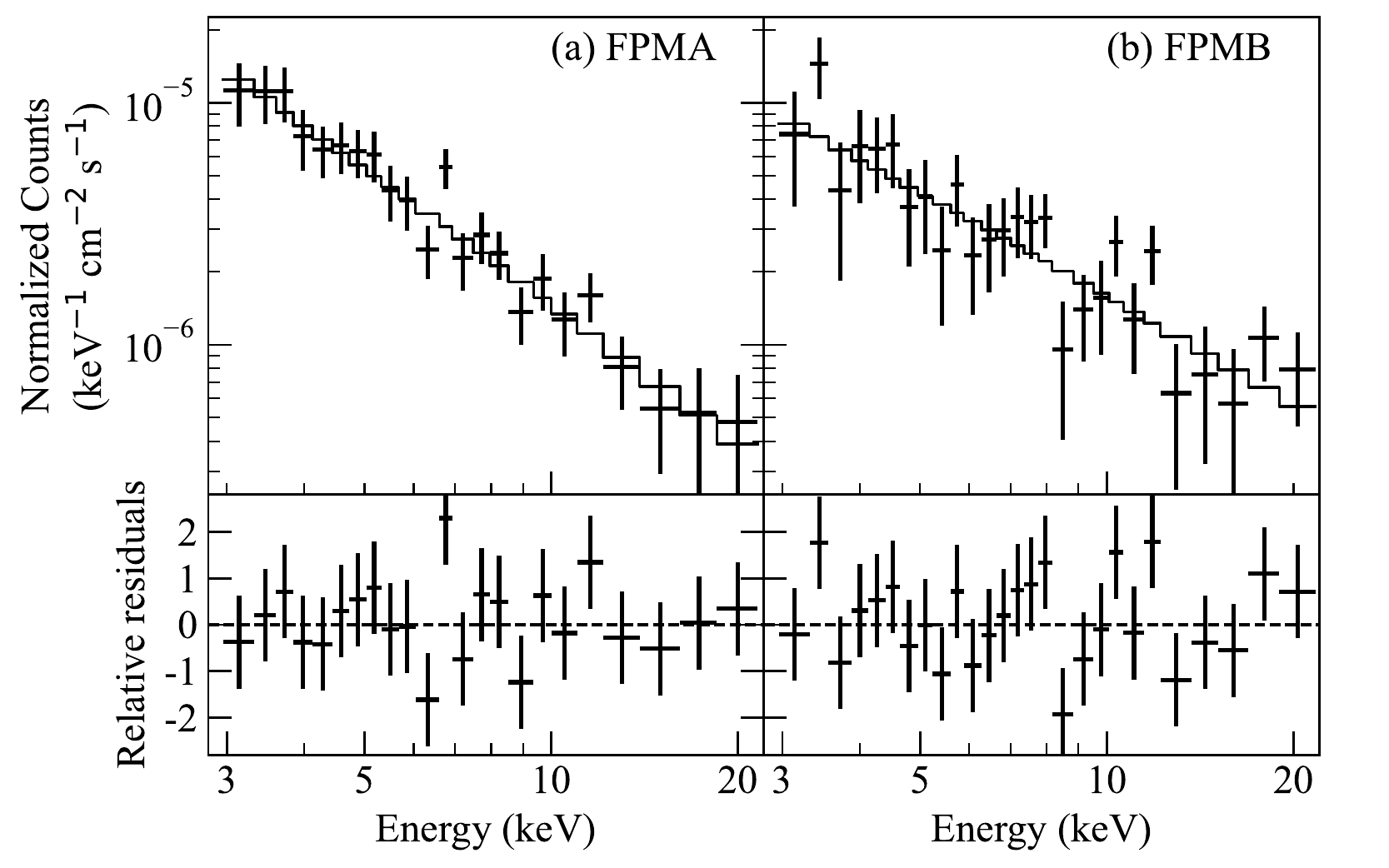}
 \end{center}
 \caption{The background subtracted spectra and best-fit model are shown as the crosses and solid line,respectively. The right and left panels indicate FPMA and FPMB, respectively.}\label{fit_spec}
\end{figure}

\begin{figure}
 \begin{center}
  \includegraphics[width=80mm,bb=0 0 230 230]{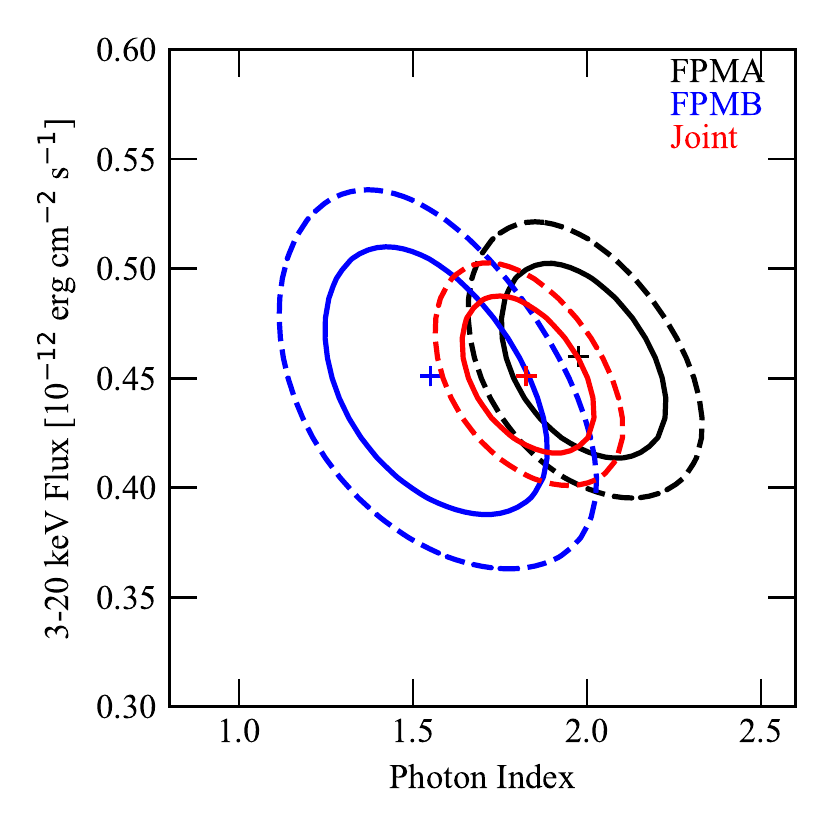}
 \end{center}
 \caption{The confidence contours for the 3-20 keV flux and photon index of the FPMA(black), FPMB(blue) and FPMA and FPMB jointed(Red) spectrum fitting. The solid and dashed lines indicate 68\% and 90\% confidence level, respectively. The crosses indicate best-fit parameters.}\label{fig:contur}
\end{figure}

We fitted the background-subtracted spectra with the absorbed power-law model as same as for \textit{XMM-Newton} and \textit{Chandra} spectra.
Figure \ref{fit_spec} shows the background-subtracted spectra and the best-fit models.
The best-fit parameters are summarized in Table \ref{tbl:Fit_Par}.
The models well reproduce the spectra with $\chi^2$/d.o.f. = 15.3/21 and 23.5/23 for FPMA and FPMB, respectively, although the FPMA spectrum prefers a slightly softer photon index of $\Gamma = 2.0 \pm 0.3$ to that of FPMB of $\Gamma = 1.6 \pm 0.4$.
To investigate the spectral difference, we calculated the confidence contours between the flux and photon index, as shown in Figure \ref{fig:contur}.  The parameters are consistent with each other to within a 68\% confidence level.
Then, we performed a joint fitting.
The best-fit parameters and confidence contours are shown in Table \ref{tbl:Fit_Par} and Figure \ref{fig:contur}, respectively.
We confirmed that the joint fitting well reproduces the FPMA and FPMB spectrum with $\chi^2$/d.o.f. = 42.3/44, and the best-fit parameters are consistent with those derived from individual fittings.

\subsubsection{Wide-band X-ray spectra}
\begin{figure}
 \begin{center}
  \includegraphics[width=\linewidth,bb=0 0 600 450]{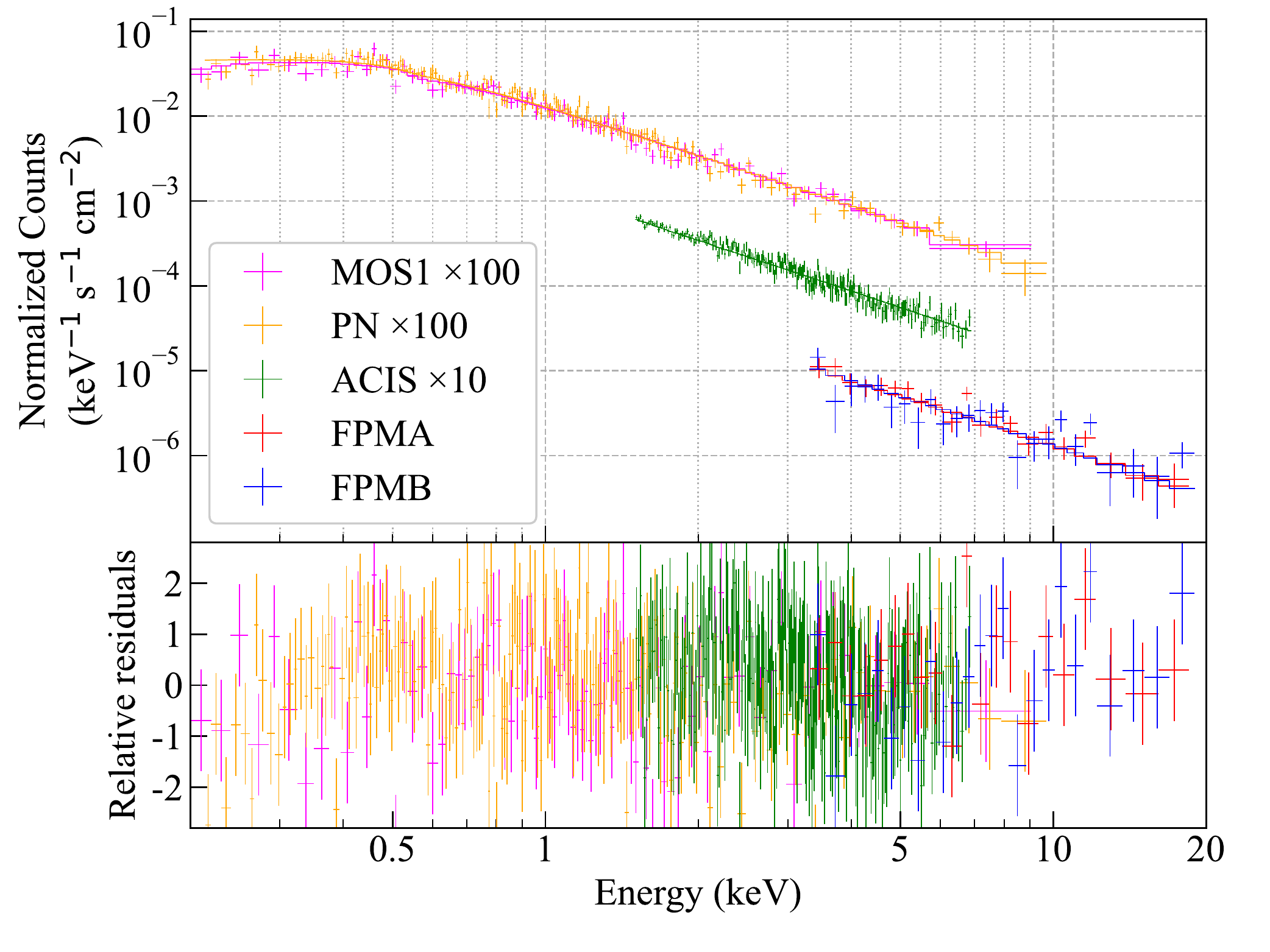}
 \end{center}
 \caption{Spectrum, best-fit absorbed power-law model and relative residuals of FPMA(red), FPMB(blue), ACIS(green), MOS1(magenta) and PN(orange) are shown. For only visibility, the counts of MOS1 and PN are 100 times scaled and counts of ACIS are 10 times scaled.}\label{fig:src_spec_all}
\end{figure}

Figure \ref{fig:src_spec_all} shows the hot spot spectra obtained by \textit{NuSTAR}, \textit{XMM-Newton} and \textit{Chandra}.
The spectra are connected smoothly, as expected from Table \ref{tbl:Fit_Par} showing that the best-fit parameters are almost consistent among different satellites within the calibration uncertainties \citep{madsen17}.
Therefore, we performed a simultaneous fitting using all the data.
The data cover a broader energy range (0.2--20\,keV) and have higher photon statistics than ever for this object.
We simultaneously fitted the spectra by the absorbed power-law model with the fixed column density.
The best-fit model well reproduced the spectrum with $\chi^2$/d.o.f.=577/551, as shown in Figure \ref{fig:src_spec_all}.
The derived best-fit parameters, summarized in Table \ref{tbl:Fit_Par}, are consistent with the parameters derived from the individual spectrum.

In order to improve the fitting, we fitted the data with the same model by allowing the hydrogen column density to vary.
The best-fit model eliminated slight residuals in 0.2--0.3 keV.
The derived parameters are shown in Table \ref{tbl:Fit_Par}.
This model gives $\chi^2/$d.o.f.=563/550, indicating a statistically significant improvement with a null hypothesis probability of $2\times 10^{-4}$ calculated by the \textit{F}-test.
The derived column density of $N_\mathrm{H}$=(4.6$\pm0.5$)$\times 10^{20}$ atoms cm$^{-2}$ is slightly higher than the Galactic value from \citet{hi4pi16}.
In contrast, the others are statistically consistent with the case of the fixed column density.
We see no spectral break in the wide-band spectra from which we excluded \textit{Chandra} spectrum below 1.5 keV.

As an advanced investigation, we searched for a spectral curvature.
If the X-ray spectrum originates from an electron synchrotron emission, a high-energy exponential cut-off is naturally expected to correspond to the maximum energy of electrons.
We examined the spectrum with a simple cut-off power-law model of \texttt{tbabs*cutoffpl}, and obtained the cut-off energy at $> 54$ keV in the statistical significance level of 5\% with $\chi^2/\mathrm{d.o.f.}=608/552$.
Since the energy of 54\,keV is above the energy coverage of our data (0.2--20\,keV).
Hereafter, we adopt 20\,keV as our conservative lower limit of the cut-off energy.

It is interesting to note the analogy with studies of SNRs.
Although X-ray synchrotron spectra in SNRs look similar to that observed in the Pictor A's western hot spot, they are interpreted differently --- SNR's X-ray are generally believed to be tails of power-law emission with an exponential cut-off below $\sim$1\,keV (see reviews, \cite{helder12,vink12} and reference therein).
For a recent example, in Kepler's SNR, \citet{nagayoshi21} revealed a power-law spectrum up to 30 keV, similar to the hot spot, and showed that it is reproduced by a power-law with an exponential cutoff at $\sim$0.5\,keV.  
Therefore, one may think that the same interpretation could be applicable to the hot spot.  However, the X-ray spectrum cannot be represented by the model of \texttt{tbabs*srcut}, which is the simple cut-off power-law model often used for SNRs if we assume cut-off energy below 20\,keV.  Hence, the cut-off energy would be higher than 20 keV for this model, too.

\begin{table*}
  \tbl{Best-fit parameters and 90\% confidence error of each detector spectrum.}
      {
        \begin{tabular}{lcccccc}
          \hline
          Instruments & Fitting range$^\mathrm{a}$ & Total Flux$^\mathrm{b}$ & Common Flux$^\mathrm{c}$ & $\mathit{\Gamma}^\mathrm{d}$ & ${N_\mathrm{H}}^\mathrm{e}$ & $\chi^2/\mathrm{d.o.f.}$ \\
          \hline
            \textit{XMM-Newton}& 0.2-10 keV\\
            MOS1    & & 0.86$\pm$0.03 & 0.18$\pm$0.01& 2.02$\pm$0.05 & 3.6 (fixed) & 97/90 \\
            PN      & & 0.89$\pm$0.02 & 0.181$\pm$0.009& 2.05$\pm$0.03 & 3.6 (fixed) & 197/181 \\
            MOS1+PN & & 0.88$\pm$0.02 & 0.181$\pm$0.007 & 2.04$\pm$0.03& 3.6 (fixed) &290/268 \\
            \hline
            \textit{Chandra}& 1.5-7.0 keV\\
            ACIS & & 0.347$\pm$0.007& 0.185$\pm$0.006 & 2.09$\pm$0.04 & 3.6 (fixed) &226/235 \\   
            \hline
            \textit{NuSTAR}& 3-20 keV\\
            FPMA   & & 0.46$\pm$0.05 & 0.20$\pm$0.03 & 2.0$\pm$0.3 & 3.6 (fixed) &15.0/21\\
            FPMB   & & 0.45$\pm$0.06 & 0.16$\pm$0.03 & 1.6$\pm$0.4 & 3.6 (fixed) & 23.5/23\\
            FPMA+B & & 0.45$\pm$0.04 & 0.18$\pm$0.02& 1.8$\pm$0.2 & 3.6 (fixed) & 42.3/44\\
            \hline
            All & 0.2-20 keV & 1.05$\pm$0.01& 0.189$\pm$0.004 & 2.02$\pm$0.02& 3.6 (fixed) &577/551\\
             & 0.2-20 keV & 1.07$\pm$0.02& 0.185$\pm$0.004 & 2.07$\pm$0.03& 4.6$^{+0.5}_{-0.5}$ &563/550\\
            \hline
        \end{tabular}
      }
      \label{tbl:Fit_Par}
    \begin{tabnote}
        \noindent
        \hbox to6pt{\footnotemark[$a$]\hss}\unskip:The energy range used for the fitting.\\
        \hbox to6pt{\footnotemark[$b$]\hss}\unskip:Unabsorbed flux within fitting range in the unit of $10^{-12}\,$ erg s$^{-1}$ cm$^{-2}$.\\
        \hbox to6pt{\footnotemark[$c$]\hss}\unskip:Unabsorbed flux within 3-7 keV in the unit of $10^{-12}\,$ erg s$^{-1}$ cm$^{-2}$.\\
        \hbox to6pt{\footnotemark[$d$]\hss}\unskip: Photon Index of power-law model.\\
        \hbox to6pt{\footnotemark[$e$]\hss}\unskip: Hydrogen column density in unit of $10^{20}$ atoms cm$^{-2}$.
     \end{tabnote}
\end{table*}

\section{Discussion}
\subsection{Origin of the X-ray emission}
The X-ray emission mechanism of the Pictor A western hot spot has been discussed since the first clear detection by \textit{Chandra} in \citet{wilson01}.
The X-ray spectrum is brighter and harder than the extrapolation of the radio to optical synchrotron spectrum.
The SSC process is known to be challenging to explain the X-ray because the X-ray and synchrotron radio emissions have different spectral indices and emission regions \citep{wilson01,thimmappa20}.
It is considered plausible that the X-ray originates from a synchrotron emission of the electrons different from that corresponding to the radio to optical emission (see e.g., \cite{wilson01,hardcastle04,hardcastle16,thimmappa20}).

\citet{tingay08} proposed that X-rays originate from the fine structures around the radio brightness peak, and \citet{hardcastle16} supported the scenario based on the time scale of the flux decrease.
However, according to \citet{hardcastle16} and \citet{thimmappa20}, a major part of the X-ray emission comes from the away of the radio fine structure.
In addition, as has been checked in Section \ref{sec:chandra_time_val}, an instrumental effect (i.e., contamination) may contribute the observed flux drop, especially on the soft energy band, and thus the intrinsic variability may be reduced. 
Therefore, as pointed by \citet{thimmappa20}, most of X-rays emitted from the thin shock front region seems to be more reasonable.

\subsection{Accelerated electron energy distribution}
Hot spots are believed to be the site of the particle acceleration via the relativistic shock and the possible ultra-high-energy cosmic-ray accelerator \citep{hillas84,kotera11}.
Therefore the observational electron spectrum of the hot spot potentially provides us with essential information for the relativistic shock study.

The X-ray observations in a wide energy range with high statistics revealed the featureless power-law spectrum in 0.2--20 keV.
The spectrum indicates that the X-ray emitting electrons are not responsible for the higher end of energy distribution (see Section 3).
Therefore we estimated the X-ray emitting electron's energy index ($p$), which is the important information of the acceleration process, from the X-ray photon index ($\mathit{\Gamma}=2.07\pm0.03$) to be $p=2\mathit{\Gamma}-1=3.14\pm0.06$.

To investigate the energy index of accelerated electrons, we should consider a synchrotron radiative cooling effect.
If the cooling is effective in the observed energy range, the derived energy index of $p$ is changed from the accelerated electron's one of $p_\mathrm{acc}$ as $p=p_\mathrm{acc}+1$ (see, \cite{meisenheimer89,carilli91}).
When a radiative lifetime in the energy band is shorter than a dynamical time scale, the cooling becomes effective (see, \cite{inoue96,kino04}).
The lifetime ($\tau_\mathrm{syn}$) in a photon energy of ($E_\mathrm{ph}$) is given from the magnetic field strength ($B$) as below
\begin{equation}
    \tau_\mathrm{syn} \simeq 15 \times \left(\frac{E_\mathrm{ph}}{2\ \mathrm{keV}} \right)^{-\frac{1}{2}} \left(\frac{B}{300 \ \mathrm{\mu G}} \right)^{-\frac{3}{2}} \mathrm{year}.
\end{equation}
Here, we adopted the representative photon energy of $2$ keV and the magnetic field of $B=300 \ \mathrm{\mu G}$ calculated from the radio to optical synchrotron spectrum under the assumption of the minimum or equipartition energy condition (\cite{meisenheimer89,isobe17}).
On the other hand, the dynamical time scale must be longer than the light crossing time of $L/c\sim 3000$ yr, where $L$ is the hot spot size of $\sim$ 1\,kpc \citep{thomson95,perley97,tingay08} and $c$ is the speed of light.
Therefore, the cooling is almost undoubtedly effective in the X-ray band, and the energy index of the accelerated electron is estimated as $p_\mathrm{acc}=p-1=2.14\pm0.06$.
The index is close to the theoretical prediction in the diffusive shock acceleration under the strong schok condition of $p_\mathrm{acc}=2$ or relativistic shock of $p_\mathrm{acc} \sim2.4$ (see, e.g., \cite{bell78,spitkovsky08}).

Several authors (e.g., \cite{wilson01,aharonian02}) already pointed out the same electron energy index of the hot spot.
However, the low statistics due to only observation by \textit{Chandra}, allowed the various X-ray spectral shape.
Our study tightly determined the X-ray spectral shape and excluded the cut-off or cooling break feature in the X-ray band.

In addition to the spectral index, the maximum electron energy ($E_\mathrm{e,max}$) is observationally estimated from the synchrotron cut-off energy ($E_\mathrm{cut}$) as,
\begin{equation}
    E_\mathrm{e,max} \simeq 40 \times \left(\frac{E_\mathrm{cut}}{20 \ \mathrm{keV}} \right)^{\frac{1}{2}} \left(\frac{B}{300 \ \mathrm{\mu G}} \right)^{-1} \mathrm{TeV}.
\end{equation}
From our study, there is a lower limit of the cut-off energy as $E_\mathrm{cut}>20$ keV (see Section 3), which requires the lower limit for the maximum energy of the accelerated electrons $E_\mathrm{e,max}>40$ TeV.

We compared the maximum electron energy to blazars, which have spectral and environmental similarities with hot spots. 
Some observational and analytical studies estimated the maximum electron energy to be up to only a few TeV even in the TeV gamma-ray emitting blazars, which are believed to have the highest value among some types of blazar (see e.g., \cite{inoue96,kataoka99,kino02,MAGIC20}).
Therefore the synchrotron X-ray in the hot spots may reflect the highest energy electron in the AGN jet system.

For further investigations, we need to detect a higher energy spectrum and or other hot spots than in this study.
Due to both lower X-ray flux and smaller angular offset from the nucleus in most hot spots than in the Pictor A western hot spot, it may be difficult for \textit{NuSTAR} to detect other hot spots.
We anticipate that future hard X-ray missions, which have better angular resolution and  sensitivity than those of \textit{NuSTAR}, for example, \textit{FORCE} \citep{FORCE18} will vigorously promote this kind of study.

\section*{Acknowledgment}
We thank Mr. Yuki Imai for help in the \textit{NuSTAR} data analysis and Dr. Naoki Isobe for valuable comments.
We deeply appreciate Observational Astrophysics Institute at Saitama University for supporting the research fund.

\bibliography{reference}{}
\bibliographystyle{aasjournal}

\end{document}